\documentclass{appolb}
\usepackage{graphicx}
\usepackage{lineno}
\usepackage{cite}
\usepackage{xcolor}
\begin{document}

\newcommand{\Ncoll}{$N_\mathrm{coll}$}
\newcommand{\Npart}{$N_\mathrm{part}$}
\newcommand{\pt}{$p_\mathrm{T}$}
\newcommand{\Raa}{$R_\mathrm{AA}$}

\title{System size dependence of particle production and collectivity from the STAR experiment at RHIC%
\thanks{Presented at Quark Matter 2022}%
}
\author{Tong Liu (for the STAR collaboration)
\address{Wright Laboratory, Yale University, New Haven, Connecticut 06520}
}
\maketitle
\begin{abstract}
Medium modification of particle spectra and the origin of collectivity in small collision systems are widely debated topics in the heavy-ion community. To address these open questions, we propose the study of particle production and collectivity for varying system sizes,Au+Au $>$ Ru+Ru/Zr+Zr $>$ Cu+Cu $>$ d+Au $>$ $\gamma$+Au, available at RHIC using the STAR detector. 

We present the first measurements of centrality dependent charged hadron production in Isobar (Ru+Ru and Zr+Zr) collisions,
including the nuclear modification factors ($R_\mathrm{AA}$) at high transverse momentom ($p_\mathrm{T}$), and identified particle spectra at low $p_\mathrm{T}$\ at mid-rapidity. 
Combined with existing results from other systems, they probe system size and collision geometry dependences of the medium modification to particle production.

In addition, we present the measurement of particle production and long-range di-hadron correlations in $\gamma$+Au events using ultra-peripheral Au+Au collisions at RHIC.

\end{abstract}
  
\section{Introduction}
Since the discovery of the quark-gluon plasma (QGP)\cite{Busza_2018}, its various properties and how they depend on initial conditions have been under active investigation. One such effort is the beam energy scan (BES) performed by RHIC\cite{STAR:2010vob}, which studies how QGP properties change with the collision energy. Yet, there is another equally important dimension, i.e. the system size dependence.
RHIC provides a large variety of collision systems, covering the number of participating nucleons (\Npart) from 1 to a few hundred. 
Furthermore, systems with similar \Npart\ but different initial geometry can be used to study how medium properties evolve with the collision system.

In these proceedings, we present the following results from the STAR experiment.
In Sec.\ref{sec:isobar_highpt}, we first focus on the high-\pt\ charged hadron spectra at $\sqrt{s_{NN}}=$ 200 GeV and compare their \Raa\ in different systems, namely Au+Au, Isobar, Cu+Cu and $d$+Au. In Sec.\ref{sec:isobar_lowpt}, we focus on the 200 GeV Isobar collision system and explore the low-\pt\ spectra of identified charged hadrons, especially the yield ratios between Ru+Ru and Zr+Zr. 
Finally, in Sec.\ref{sec:photonuclear}, we study photonuclear events tagged in 54 GeV Au+Au collisions, and present a study probing the long-range correlation in such events. 

\section{Nuclear modification of hard partons in Isobar collisions}
\label{sec:isobar_highpt}
In this section and Sec.\ref{sec:isobar_lowpt}, we utilize the isobaric collision data taken by STAR in 2018. A detailed description of the run, event, and centrality selection can be found in Ref.\cite{star_cme}. 
The efficiency corrected mid-rapidity ($|\eta|<0.5$) charged particle spectra are divided by the $p$+$p$ spectrum scaled by the number of binary collisions (\Ncoll) from Ref.\cite{star_auau200_hadron} to obtain the \Raa.  

\begin{figure}[htb]
\includegraphics[width=12cm]{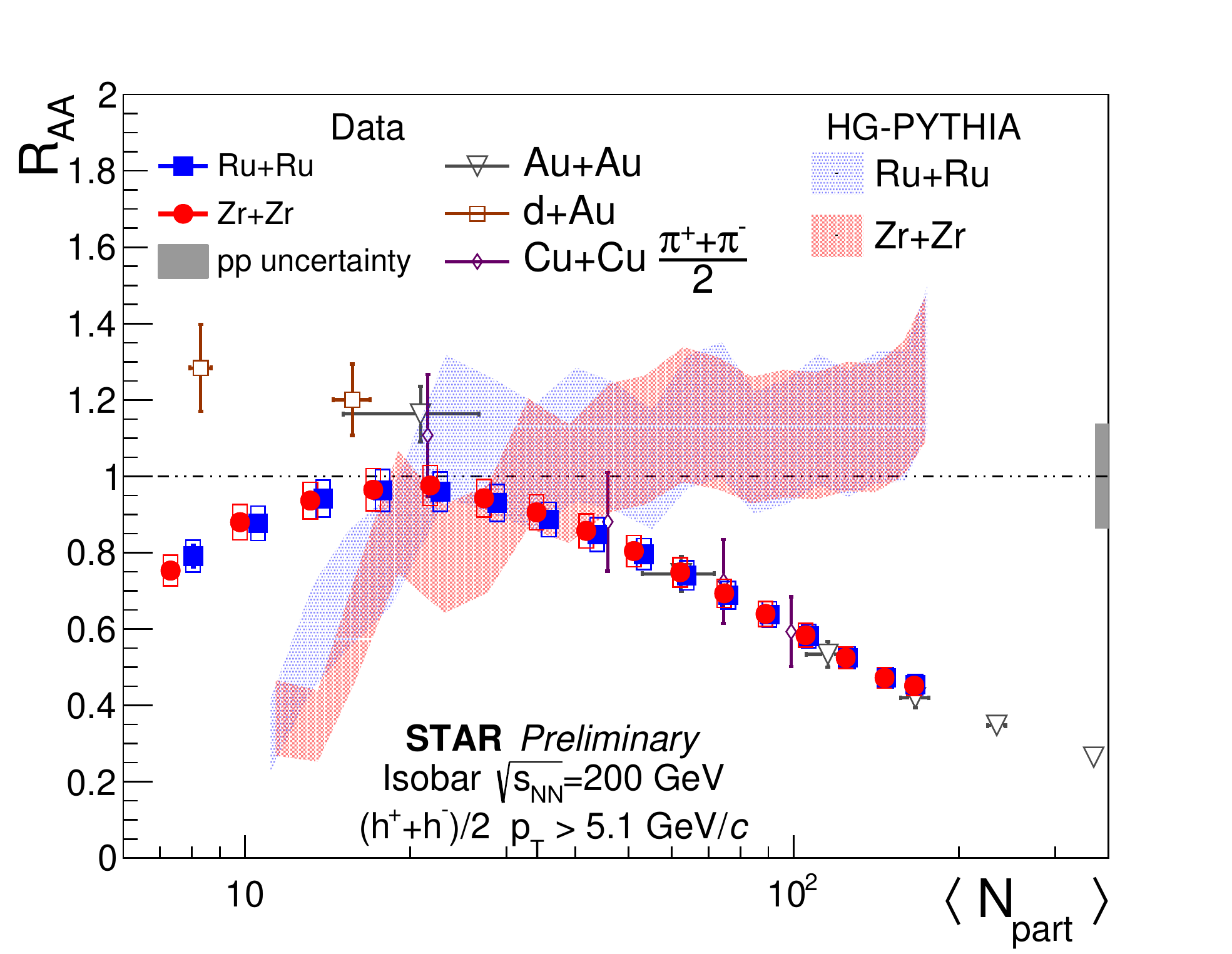}\hspace{2pc}%
    \caption{
    \Raa\ of inclusive charged hadrons with \pt $>$ 5.1 GeV/$c$. For the isobar results, each point represents 5\% centrality (rightmost: 0-5\%; leftmost: 75-80\%). The $p$+$p$ and Au+Au spectra are taken from \cite{star_auau200_hadron}, d+Au from \cite{STAR_dAu_hadron}, Cu+Cu from \cite{STAR_CuCu_pion}. HG-PYTHIA simulations modified from \cite{hg_pythia_Loizides_2017} are shown as shaded bands. See text for detail.}
    \label{fig:highpt_raa} 
\end{figure}

In order to study the dependence of the medium modification of high-\pt\ charged hadrons on system size, \Raa\ for charged hadrons integrated above 5.1 GeV/$c$ is plotted as a function of \Npart. Thanks to the high statistics of the Isobar dataset, we perform this analysis in 5\%-wide centrality bin.
Shown in Fig.\ref{fig:highpt_raa}, \Raa\ from Isobar collisions increases gradually from 0-5\% central to 55-60\% peripheral collisions, which aligns with the expectation that hard partons experience less quenching as \Npart\ decreases. When compared against similar measurements in other collision systems (\cite{star_auau200_hadron,STAR_dAu_hadron,STAR_CuCu_pion}), we observe that in this region, similar \Npart\ leads to similar \Raa\  regardless of the collision system. This corroborates that \Npart\ is the dominating factor in modification of hard partons.
%, at least on the event level. 
The Ru+Ru and Zr+Zr \Raa\ results are consistent with each other within uncertainties.

In more peripheral events, one would naively expect smaller quenching effects and a larger \Raa. Actually, as cold nuclear matter effects become dominant, 
\Raa\ in smaller systems
(i.e. $p$+Pb and $d$+Au) 
has been observed to be consistent with or even above one at intermediate \pt \cite{STAR_dAu_hadron, ALICE_pPb:2014nqx}.
However, the \Raa\ values in Isobar collisions with \Npart $<$ 20 start to decrease from 60\% up to 80\% peripheral events, and deviate from previous small system measurements. Event selection and geometry biases may contribute to this deviation, as demonstrated by the HG-PYTHIA \cite{hg_pythia_Loizides_2017}, which shows that the number of hard scatterings per nucleon-nucleon collision is smaller in peripheral events, and the effect is significantly stronger when selecting centrality via experimental observables compared to via the impact parameter. This effect is also observed by CMS\cite{CMS_hgpythia_PhysRevLett.127.102002} and ALICE\cite{ALICE_peri_bias_PhysRevC.91.064905}. Here we show simulations for the Isobar data using HG-PYTHIA. To maximally mimic the experimental condition, we use the number of charged particles with $|\eta|<0.5$ and \pt\   $>$ 0.2 GeV/$c$ as the centrality indicator, calculate the invariant yield for charged particles with \pt\ $>$ 5 GeV/$c$, and compare to simulated $p$+$p$ yield as in real data. 
The results (shaded bands in Fig. \ref{fig:highpt_raa}) qualitatively predict the decrease in \Raa\ at similar \Npart\ as that observed in data; yet quantitatively it over-predicts the bias.
A more accurate prediction needs more detailed simulation, e.g. inclusion of detector effects, to see if quantitative agreements can be reached.

\section{\pt\ distribution of identified particles in Isobar collisions}
\label{sec:isobar_lowpt}

While high-\pt\ particle yields can indicate hard parton modification in the medium, bulk properties of the produced medium, e.g. flow velocity, are usually extracted via soft hadron spectra.
Because of the well-controlled and similar running conditions of the two Isobar species, detector effects and systematic uncertainties largely cancel when taking the yield ratios between them.
Therefore, their raw yield ratio can already provide ample physics information.

\begin{figure}[htb]
\includegraphics[width=12.5cm]{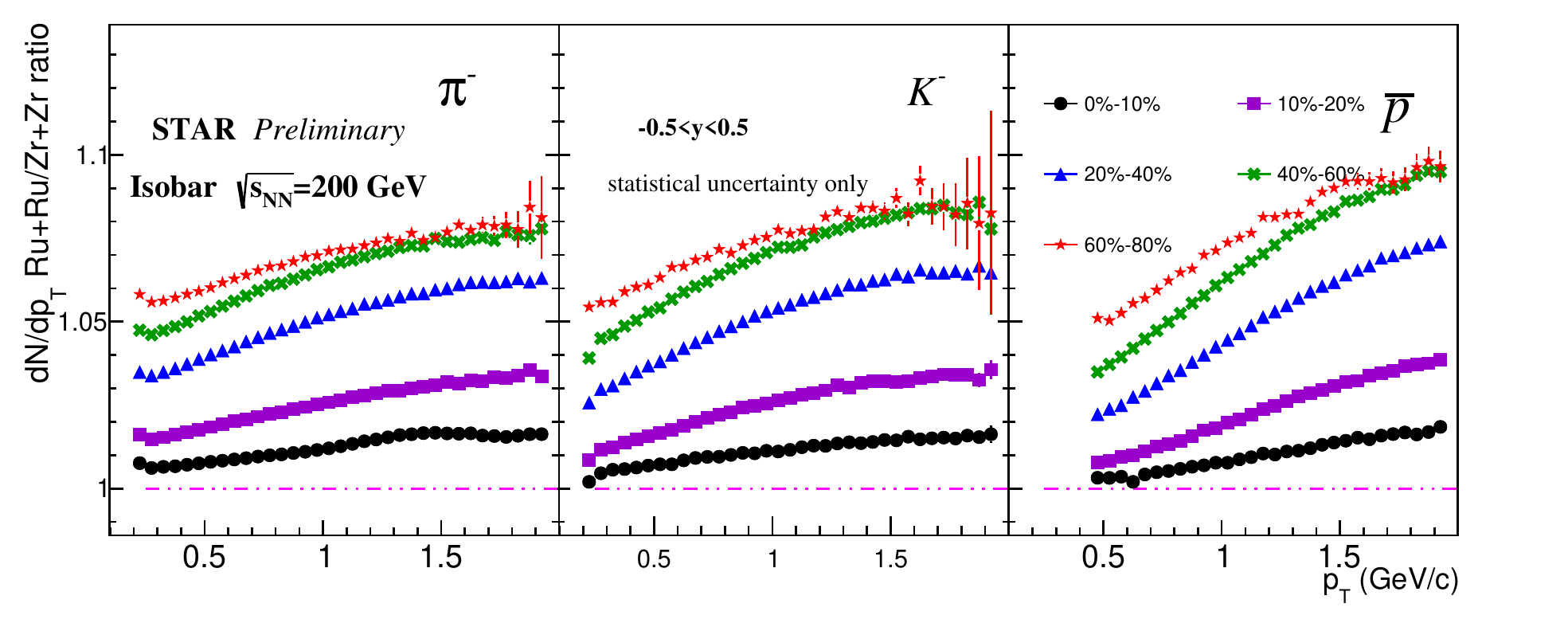}\hspace{2pc}%
    \caption{ 
    Mid-rapidity yield ratios of ${\pi}^-$, $K^-$ and $\overline{p}$ between Ru+Ru and Zr+Zr collisions as a function of \pt, in different centrality classes.  Ratios of positive particles show similar trends.}
    \label{fig:pid_ratio_plot}
\end{figure}

We use the same event sample and centrality selection as in Sec.\ref{sec:isobar_highpt}, and follow the particle identification method outlined in Ref.\cite{PID_method_PhysRevC.88.014902} to extract identified particle spectra for $|y|<0.5$. 
As seen in Fig.\ref{fig:pid_ratio_plot}, ratios of all species show a centrality dependence, with the most central events closest to 1 and more peripheral events moving up gradually; we also observe the ratio increases with \pt. This is consistent with the Glauber model, which predicts larger \Npart\ and \Ncoll\ ratios between Ru+Ru and Zr+Zr with increasing centrality\cite{star_cme}.
On the other hand, the Ru+Ru/Zr+Zr ratio of each species at a given centrality has subtle differences across \pt. Namely, the slope for protons is larger than that for $K$ mesons and subsequently $\pi$ mesons; the relationship is similar across centralities. 
This indicates the trend is driven by the difference in the radial flow between Ru+Ru and Zr+Zr collisions.

\section{Long-range correlations in photonuclear processes}
\label{sec:photonuclear}

\begin{figure}[htb]
\includegraphics[width=12cm]{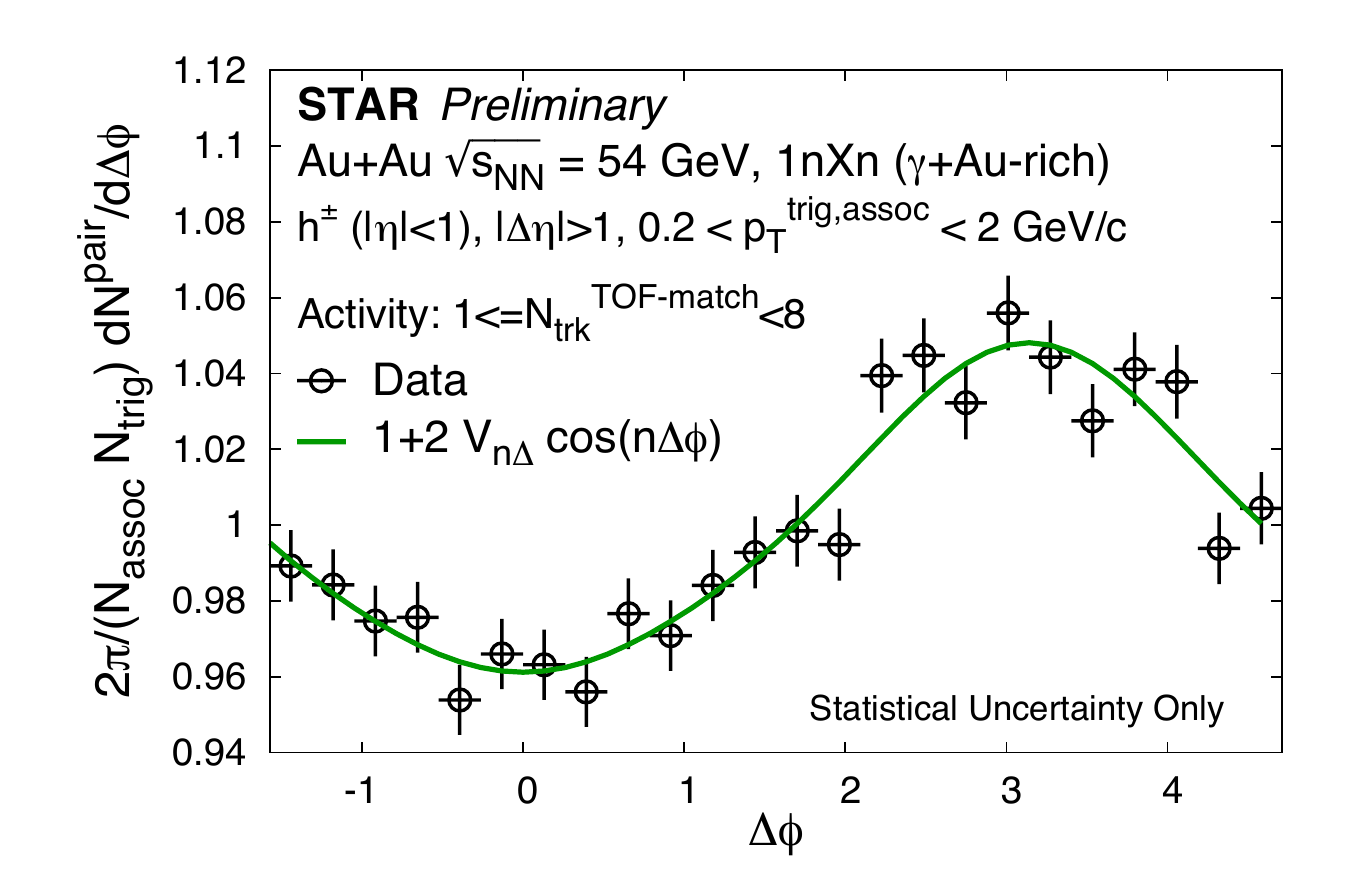}\hspace{2pc}%
    \caption{
    Long-range azimuthal correlation of charged particles produced in photonuclear processes tagged in 54.4 GeV ultra-peripheral Au+Au collisions.
    }
    \label{fig:photonuclear} 
\end{figure}

A system size scan of QGP signatures would not be considered complete without pushing our limits towards smaller systems. In this section, we present measurements of long-range  correlations in tagged $\sqrt{s_\mathrm{NN}} =$ 54.4 GeV ultra-peripheral Au+Au collisions, in search for signatures of collectivity in photonuclear processes. The tagging procedure is applied to select a $\gamma$+Au rich sample, where one nucleus only takes part in the collision by emitting a virtual photon interacting with the other nucleus, and is achieved by taking advantage of the asymmetric nature of the $\gamma$+Au processes ( \cite{Brandenburg:2022hrp,qm22_nicole}).

In events where the number of tracks detected by the Time Projection Chamber (TPC) and matched to the Time of Flight (TOF) detector falls in the range of $1<=N_{\rm trk}^{\rm TOF}<8$, we measure the two-dimensional correlation function in relative azimuthal angle $\Delta\phi$ and relative pseudorapidity $\Delta \eta$ of particle pairs. We select TOF-matched tracks as ``trigger particles'' and associate each trigger particle with the remaining tracks with $|\eta|<$1 and $0.2<p_{T}^{\rm trig,\,assoc}<2$ GeV/c. The number of ``trigger particles'' is denoted as $N_{\rm trig}$. We obtain the correlation function in $\Delta\phi$, integrated over $|\Delta\eta|>1$, and perform a Fourier decomposition as
\begin{equation}
 \label{eqn:long_range_correlation}
Y(\Delta \phi, |\Delta|\eta>1) \equiv \frac{2\pi}{N_{\rm trig} N_{\rm assoc}} \frac{dN^{\rm pair}}{d\Delta\phi} = 1+ \sum_n 2 V_{n \Delta} \cos (n\Delta \phi),
\end{equation}
 where $V_{n\Delta}$ are the Fourier coefficients, $n$ is the order of harmonics, and $N_{\rm assoc}$ is the average number of pairs per trigger particle. Figure \ref{fig:photonuclear} shows the normalized yield $Y(\Delta \phi)$ and the decomposition. No ridge-like component, i.e., a significant $Y(\Delta\phi)$ enhancement near $\Delta\phi=0$ that is considered the signature of collectivity, is seen within 
 uncertainties.
 We aim to extend these measurements with high statistics $\gamma$+Au-rich event samples using planned 2023 and 2025 data of Au+Au collisions at $\sqrt{s_{\mathrm{NN}}}=200$ GeV, and the extended $\eta$ coverage offered by the STAR forward upgrade\cite{qm22_xusun}.

\section{Summary}

In these proceedings, several studies that improve our understanding on the existence and properties of the QGP in different collision systems, and their dependence on initial conditions are presented. We show via the Isobar results in combination with other measurements, that inclusive \Raa\ of high \pt\ hadrons is mostly driven by \Npart\ regardless of initial geometry. In peripheral events of the Isobar collisions, \Raa\ is observed to be lower than unity, which may be due to event selection and geometry bias in the measurement. We also observe that identified particle yield ratios between Ru+Ru and Zr+Zr systems show centrality dependence, and the slopes of these ratios as a function of \pt\ are larger for heavier particles. These results demonstrate our sensitivity to the difference between the Isobar species. We also explore potential collectivity in photonuclear processes via tagged ultra-peripheral Au+Au events. No ridge-like structure associated with collectivity is found.
These studies will help us understand how QGP signatures depend on the system size and geometry, and contribute to the ``system size scan'' which will explore the evolution of QGP properties with the initial state. 

\bibliographystyle{hrev2}
\bibliography{ref}
\end{document}